# Status of the High Field Cable Test Facility at Fermilab

G. V. Velev, D. Arbelaez, C. Arcola, R. Bruce, V. Kashikhin, S. Koshelev, A. Makulski, V. Marinozzi, V. Nikolic, D. Orris, S. Prestemon, G. Sabbi, T. Tope, and X. Yuan

*Abstract*— Fermi National Accelerator Laboratory (FNAL) and Lawrence Berkeley National Laboratory (LBNL) are building a new High Field Vertical Magnet Test Facility (HFVMTF) for testing superconducting cables in high magnetic field. The background magnetic field of 15 T in the HFVMTF will be produced by a magnet provided by LBNL. The HFVMTF is jointly funded by the US DOE Offices of Science, High Energy Physics (HEP), and Fusion Energy Sciences (FES), and will serve as a superconducting cable test facility in high magnetic fields and a wide range of temperatures for HEP and FES communities. This facility will also be used to test high-field superconducting magnet models and demonstrators, including hybrid magnets, produced by the US Magnet Development Program (MDP). The paper describes the status of the facility, including construction, cryostat designs, top and lambda plates, and systems for powering, and quench protection and monitoring.

TABLE I
HFVMTF PARAMETERS

| Parameter | Value |
|---|---|
| Background Dipole Field | 15 T |
| Minimum Operational Temperature | 1.9 K |
| Maximum Magnet Diameter | 1.3 m |
| Maximum Magnet Length | 3.0 m |
| Maximum Stored Energy | 20 MJ |
| Maximum Weight | 20 t |
| Test Sample Temperature | 4.5-50.0 K |
| Maximum Test Sample Current (direct) | 16 kA |
| Maximum Test Sample Current (transformer) | 100 kA |

## I. INTRODUCTION

THE US DOE Offices of Science, High Energy Physics (HEP) and Fusion Energy Sciences (FES) programs have joined their efforts to build a High Temperature Superconductor (HTS) cable-testing facility. This new facility, called HFVMTF, is being constructed at Fermi National Accelerator Laboratory (Fermilab). For the US FES and HEP communities, it will serve as a test stand for HTS cable samples in high dipole fields and at increased temperatures. It is designed to provide similar or better capabilities than the European test stands, EDIPO at PSI and FRESCA2 at CERN, Switzerland [1,2]. For the US HEP Magnet Development Program (MDP) [3], it will be the main facility for the testing of magnets with fields exceeding 16 T, including hybrid magnets that are built using low temperature and high temperature superconductors. These magnets are an important step toward the ≥20 T dipoles for future hadron-hadron colliders. The paper describes the current status of the facility, including construction, cryostat designs, top and lambda plates, systems for powering, and quench protection and monitoring.

This facility will be built in Fermilab's Industrial Building One, close to the current Vertical Magnet Test Facility, in order to take advantage of this building's existing cryogenic, power, water, and crane infrastructure.

## II. TEST FACILITY PARAMETERS

The test facility parameters were selected after discussion with the community of users. These parameters are documented in [4].

To achieve maximum cooling, the magnet providing the background field will operate in superfluid helium at a temperature of 1.9 K. This requirement imposes the use of a lambda plate in the cryostat assembly. By design, the operational background dipole field provided by the magnet is 15 T [4,5]. The maximum stored energy in the magnet is on the order of 20 MJ. The test facility is designed to have an operational lifetime of at least 20 years. Table 1 summarizes the HFVMTF parameters.

The cryostat for the HTS test sample will be inserted into the magnet aperture. Its conceptual design is similar to the cryostat of 4.5 to 50 K. To excite HTS test samples with a current up to 100 kA, a superconducting transformer will be used, a solution that has been implemented in the SULTAN [6] and EDIPO [1] test facilities.

Manuscript received October xx, 2022. This work was supported by the US Department of Energy Offices of Science, High Energy Physics, and Fusion Energy Sciences, and the Fermi Research Alliance, LLC, under contract No. DE-AC02-07CH11359 with the US Department of Energy *(Corresponding author: G. Velev.)*

G.V. Velev, C. Arcola, R. Bruce, V. Kashikhin, S. Koshelev, A. Makulski, V. Marinozzi, V. Nikolic, D. Orris, T. Tope, and X. Yuan are with the Fermi National Accelerator Laboratory, Batavia, IL 60510, USA (e-mail: velev@fnal.gov).

D. Arbelaez, S. Prestemon, and G. Sabbi are with the Lawrence Berkeley National Laboratory, Berkeley, CA 94720, USA.

Color versions of one or more of the figures in this paper are available online at http://ieeexplore.ieee.org.

Digital Object Identifier will be inserted here upon acceptance.

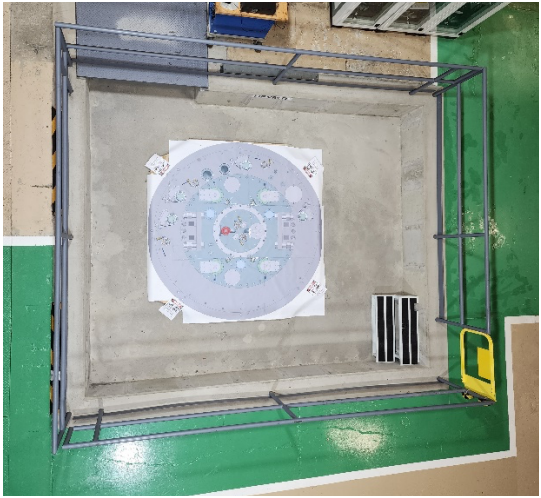

Fig. 1. Top view of the HFVMTF pit and shaft opening.

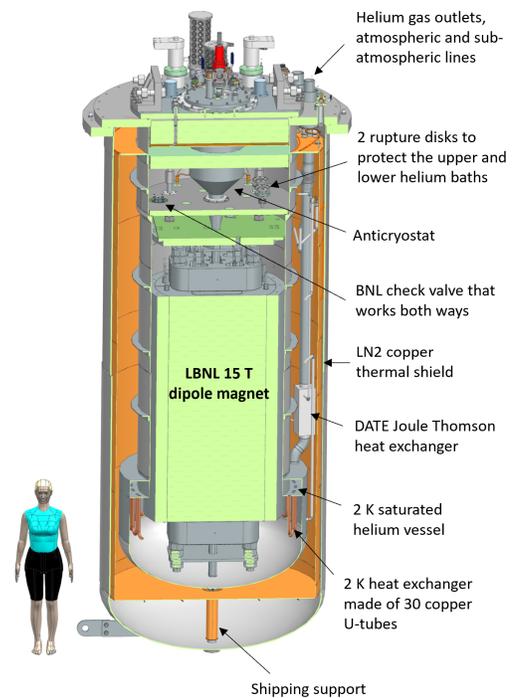

Fig. 2. Cross-section of the conceptual design of the cryostat with the magnets inside.

### III. STATUS OF PIT CONSTRUCTION

The civil construction of the test facility began in November 2020. In this part of the project, the majority of the work performed involved the excavation of a shaft in the existing building and the installation of magnetic field shielding and a fiberglass liner tube. A stainless-steel ring was added to the top edge around the shaft to achieve even load distribution of the weight of the cryostat and magnet on a concrete slab. All of the components were cast-in-place with concrete around the fiberglass liner which was used to form the pit wall area. The shaft opening in the pit will allow us to install a cryostat with a diameter of 2.18 m and a length of 6 m. At the same time, a two-trench path was constructed in the building to allow the power bus and cryogenic process piping to be connected to existing process piping, a connection that is necessary for the operation of the cryostat. The civil construction part of the project was successfully completed and closed in May 2021. Figure 1 shows the completed pit with a top view, together with a 1:1 scale print that depicts the shaft opening.

### IV. CRYOSTAT DESIGN

The HFVMTF cryostat is a large double-bath vessel with a lambda plate that separates the 4.5 K normal liquid helium on the upper section from the pressurized superfluid helium at 1.9 K and 1.2 bar [7]. The current design of this helium vessel, shown in Fig. 2, is similar to the existing VMTF cryostat [8] located in IB1 at Fermilab. The first conceptual design has been performed at Fermilab, and it is being manufactured externally following the latest ASME standards. The cross-section of the conceptual design of the cryostat with the magnets inside is shown in Fig. 2.

The entire inner helium vessel will be supported by the top flange and the maximum design pressure of this vessel will be 100 psi (6.9 bar). The pressurized superfluid bath will be able to support a 20 plus ton magnet with a maximum diameter of 1.3 m and a maximum length of 3 m.

The inner helium vessel is composed in the upper section by a horizontal lambda ring, where the lambda plate will lay. The lambda plate supports the weight of the magnet and any additional pressure loads. Differently from the other parts of this cryostat, the lambda plate must meet the requirements of ASME because the pressure energy will be contained in the inner helium vessel if one of the inside parts breaks (e.g., the heat exchangers). The lambda plate will be tested in closed conditions to simulate the worst-case scenarios with the maximum loads to guarantee the safe operation of the entire test stand. The next section of this paper details the equipment on this plate and all of the safety requirements associated with it.

Another important part of this cryostat is the saturated superfluid vessel. This vessel is composed of a ring attached to the lower part of the helium vessel and has an internal pressure of about 0.4 psi (0.03 bar) under operating conditions. This vessel also has a liquid-to-liquid heat exchanger composed of 30 copper U-tubes, allowing the pressurized superfluid helium bath from this sub-atmospheric vessel to cool. The same type of heat exchanger is already used at the magnet test facility at CERN [9,10]. The axisymmetric design allows the exchanger to maintain the magnet in a centered position so as to minimize the imbalance forces from an external magnetic shield.

The helium gas outlet of the saturated superfluid vessel is connected to a small Joule Thomson (JT) heat exchanger manufactured by DATE. This counterflow heat exchanger is designed to transfer 28 W from the hot high-pressure line (coming from the upper helium bath at 4.5 K) to the cold low-pressure line (coming from the saturated vessel at 1.9 K). This device will reduce the temperature of the liquid helium upstream of the Joule Thomson valve from 4.5 K to 3.5 K with a maximum mass flow of 3 g/s.

Finally, a copper thermal shield cooled by circulating liquid nitrogen will intercept the radiative heat load coming from the vacuum jacket. The thermal shield, the helium vessels and the pipes will also be covered with several layers of MLI to minimize the radiative heat load. The pin at the bottom of the cryostat supports the cryostat during shipping when the cryostat is oriented horizontally. This shipping support will be removed after delivery at Fermilab and will be replaced by a part composed of several parallel aluminum plates maintained by G10 rods to reduce both the conductive and radiative heat load on the superfluid bath.

## V. Lambda plate design and safety

The lambda plate acts as a thermal and pressure shield between normal liquid helium at 4.4 K and the pressurized superfluid bath at 1.9 K. The two main concerns with this cryostat are to safely operate the magnets that will be tested inside the superfluid bath and to minimize the heat load on the superfluid bath. Large pressure differences across the lambda plate can occur during the quench of the 25-ton superconducting magnet. To reduce the mechanical stress caused by the weight of the magnet and the additional pressure force, the plate has a thickness of 2 inches (50 mm) and a diameter of 58 inches (1470 mm), is made of 304L stainless steel (the same material as the cryostat) and is covered on its top with G10 composite. The G10 part acts as a thermal insulator and will be glued to the stainless steel using Stycast®. Generally, double-bath cryostats use conical surface-to-surface contact seals [3] in between the normal and superfluid helium baths. In this case, the diameter of the lambda plate is large compared to other double-bath test stands, and superfluid helium leaks will generate large heat leaks that will account for a majority of the 2 K heat leaks. To minimize the leaks at the location of the lambda ring, a spring-energized seal [11] solution has been chosen. This solution is already used by the Brookhaven National Laboratory for their 1.9 K vertical test facility and it significantly reduces superfluid heat leaks. A minimum of 8 tons is necessary to perfectly seal this plate.

To accommodate the pressure across the lambda plate during the cooling process and the quench of the magnet, the HFVMTF cryostat will use the check valve designed by BNL. This valve will be installed on the lower side of the lambda plate. The BNL valve is actually composed of two check valves. The larger one is made of Teflon that protects the superfluid bath from overpressurizing during the quench of the magnet. The second valve, called the reverse check valve, is smaller and is located inside the other valve to accommodate the pressure between the two baths during the cool down between 4.5 K and 1.9 K. The opening pressure for both check valves is 0.1 bar and the maximum pressure drop through the larger check valve is 0.4 mbar during a quench event.

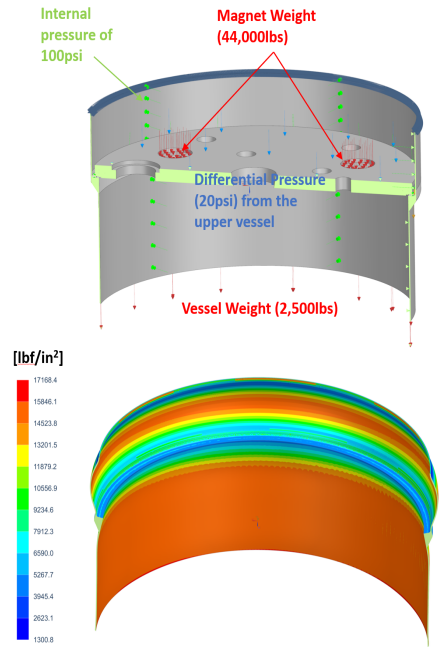

Fig. 3. FEA study of lambda plate deflection.

Two rupture disks will also be installed on the lambda plate. They will both be reverse buckling disks with a burst pressure of 15 psi (1 bar) at 4.5 K and will resist full vacuum. These disks are both located in the upper bath for better accessibility. Burst sensors will monitor the status of these two devices during cool down and after a quench of the superconducting magnet. The rupture disks are not supposed to burst during quench, but high-pressure peaks can appear in the lower vessel during this sudden event and partially damage the equipment. A 2-inch rupture disk is necessary to protect the pressurized superfluid bath during the worst-case scenario, which can occur if we lose vacuum insulation in the cryostat. This catastrophic event would almost instantly cause the quench of the superconducting magnet and a significant amount of supercritical helium, at about 5 K and 85 psi (6 bars), would be released. The helium bath above the lambda plate would be protected by a 1-inch rupture disk. This second rupture disk protects the upper bath during the cooling process between 4.5 K and 1.9 K if the reverse check valve fails to open. These two devices will add significant heat loads to the superfluid bath (about 3 W), mainly due to their small thickness (0.002-inch or 0.05 mm).

In addition to these safety devices, three 2-inch diameter support rods attached to the lambda plate will support the magnet, and similar feedthroughs to the VMTF cryostat will be used for the instrumentation wires, voltage taps, and power leads that reach the lower vessel. These feedthroughs are G10 composite conical plugs with a diameter of about 2 inches. The wires will be glued to the G10 plugs using Stycast®, and Apiezon® grease will be used in between the plugs and the lambda plate to limit helium leaks. The HFVMTF will use the same sealing solution as the VMTF cryostat for the anti-cryostat that will go through the lambda plate. A thick conical Teflon part will be installed in place of the anti-cryostat when the test stand is used as a magnet testing facility.

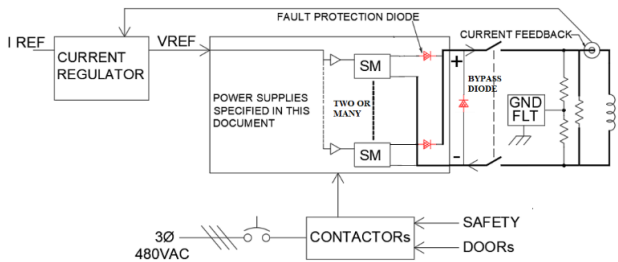

Fig. 4. Power supply system diagram.

In addition, Fermilab performed an FEA study to verify the mechanical properties of the cryostat under the worst-case conditions. The maximum calculated stress on the lambda plate and the wall of the inner helium vessel are presented in Fig. 3, when applying an internal pressure of 100 psi (6.9 bars) on the wall and a differential pressure of 20 psid (1.3 bar) from the upper vessel on the lambda plate. In this case, the maximum stress on the wall of the inner helium vessel is below 18,000 psi (124.1 MPa), less than 20,000 psi (137.9 MPa) required for dual 304/304L, and the deflection of the lambda ring is below 0.005 inches. This value is the limit that guarantees that there are no superfluid leaks in between the lower and the upper vessel at the location of the spring-energized seal.

## VI. POWER SYSTEM

A 24 kA ≥ 20 V and a 16 kA ≥ 20 V power system are specified for powering the main magnet and the magnet insert individually, or simultaneously for hybrid configurations.

It is specified that power systems be voltage-regulated. A ground-referenced external voltage will be generated and supplied from an external current loop to regulate the load current. Each system will operate with inductive loads ranging from 20 μH to 120 mH with energy extraction resistors. Using multiple switch-mode power supplies in parallel is one of the options that is preferred by the specification for its higher power efficiency, smaller size, and ease of swapping out faulty units. Possible power system topologies are a combination of master/slave units as needed to achieve the output ratings. In operation, the currents between different power supply units should be matched within 10%.

Figure 4 shows the system diagram for a switch-mode power system powering inductive loads.

It is necessary that protection diodes be connected in series and parallel with each power supply element. The series diode will protect any unit having the other parallel units drive energy from a possible internal fault into the fault. The parallel diode (the bypass diode) will be implemented by Fermilab and will protect the output of each unit from having a negative voltage greater than -1.5V applied to the output when the magnet is sourcing the current. All fault protection diodes are rated for the full output of each power supply unit. After a magnet quench is detected, both dump switches open, and the load current flows through the energy extraction resistors. The time constant of large superconductive magnet loads can be very long when the supplies are set to zero (bypass mode). The decay time will be determined by the voltage drop due to the diode and current bus resistance when the energy extraction resistors are not switched into the circuit.

TABLE 2
HFVMTF POWER SUPPLIES ELECTRICAL AND MECHANICAL SPECIFICATIONS

| Specification | Value |
| --- | --- |
| Input Voltage | 480 VAC, 60 Hz |
| Output Voltage | >20 V |
| Output Currents | 24 and 16 kA |
| Power Supply Units Cooling | Water Cooling |
| Efficiency for ≥50% Load | >80% |
| Voltage Regulation | <±0.5% |
| Voltage Regulation Upper Limit | >50 Hz |
| Voltage Ripple (from 20Hz to 300kHz) | 0.2% |
| Current imbalance among PS units | <10% |
| Footprint | <200 ft2 |
| Safety components | NRTL [12] rated |

The power systems will be interfacing to Fermilab's already-developed energy dump system, grounding system, quench management system, and other control systems. Personal safety and magnet safety are of the highest priority in the design, and the power systems and controls must be highly reliable and designed to be fail-safe. The key electrical and mechanical specifications are summarized in Table 2.

For the time being, the engineering team at Fermilab's magnet test facility has established the preliminary engineering for power system integration to the HFVMTF facility, and the bids for power systems are ongoing. In the past 30 years, the engineering team has successfully commissioned other 5 kA, 10 kA, 18 kA, and 30 kA power systems and kept them operating safely during their service life.

## VII. QUENCH PROTECTION AND MONITORING SYSTEM

The block diagram for the magnet's quench protection and monitoring (QPM) system is shown in Fig. 5. It consists of two independent symmetric branches that can simultaneously protect two superconducting magnets or a magnet and test sample, plus their superconducting bus and their high current vapor-cooled leads. When testing a hybrid magnet, the two branches of the QPM will simultaneously protect the LTS magnet and the HTS insert. When the system is used to test HTS samples for fusion energy R&D, the first branch will protect the main dipole magnet and the second branch the test sample. Depending on the mode of operation, one branch can be configured to switch off the current on the other one and vice-versa in case of a detected quench. This is to protect the non-quenching coils from exposure to large voltages induced by the collapsing current in the quenching LTS or HTS coil.

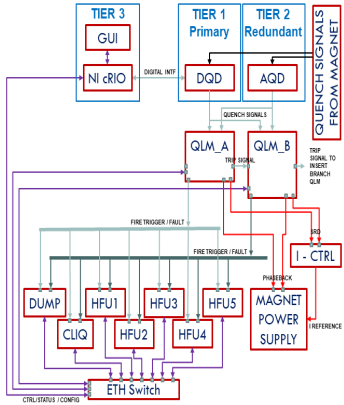

Fig. 5. A block diagram of the quench protection system.

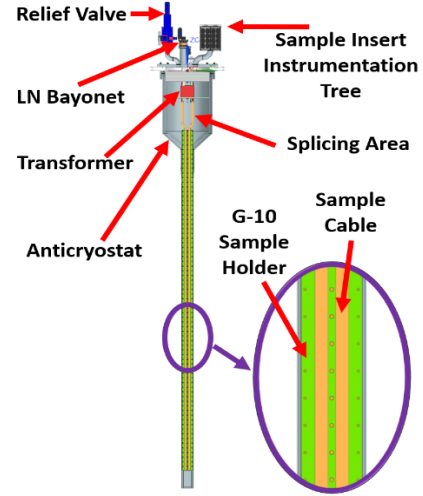

Fig. 6. Anticryostat and sample holder.

Each quench protection branch is designed to be completely independent and fully redundant, from the detectors to the energy extraction system. Each consists of both a Quench Detection and Monitoring (QDM) section and a Quench Protection (QP) section.

The QDM is based on the one developed for the Mu2e [13] experiment at Fermilab. The principle for detecting a quench in the LTS magnet and HTS insert is based on monitoring the resistive voltage growth and comparing this signal to a predefined threshold [9]. When the quench threshold is exceeded for longer than the validation time, a quench trigger is generated. The QDM section consists of a primary Digital Quench Detection (DQD) hardware system (Tier-1), a redundant Analog Quench Detection (AQD) hardware system (Tier-2), and a quench management system based on National Instruments' CompactRIO [14] system (Tier-3). The DQD provides both quench detection and quench characterization capability. Both DQD and AQD have built-in high voltage isolation, user-programmable gains and attenuations, user-configured current dependent thresholding, and validation times. The quench management system (Tier-3) provides quench configuration, control, and monitoring and quench data management for post-quench analysis.

The Quench Protection (QP) section includes a hardware Quench Logic Module (QLM) and the actual protection devices, including the Heater Firing Units (HFUs), the Coupling-Loss Induced Quench (CLIQ) units [15], and the energy extraction system (dump resistor circuit).

The QLM consists of a redundant dual FPGA board solution that carries out the critical hardware-based quench logic, protection heater-control logic, CLIQ control logic, energy extraction system enabling and discharge control, power system enabling, phase-back (PB), and slow ramp-down. Depending on which quench event trigger is generated by the detectors, the QLM initiates the quench protection logic, which results in energy extraction (dump resistor circuit), protection heater discharge, and CLIQ unit discharge with a user-specified delay from 0 to 1000 ms, individually configurable for each type of protection device.

The HFVMTF QPM system is capable of monitoring 128 isolated quench characterization channels using fast-data logers, and uses hardwired fail-safe 5 kHz signals for the PB, SRD, quench event triggers, and discharge triggers. It may control 8 single-channel HFUs for up to 32 magnet-protection heaters, two dual-channel HFUs for magnet spot heaters, and two CLIQ units. A standardized ethernet interface is used to propagate the system reset, select the mode of operation (MDP or FES), select the HFUs' charge voltage and cap bank configuration, and query the status of the QLMs and the protection devices from the Tier-3 system hosting a graphical user interface.

## VIII. Anticryostat and sample holder

The anticryostat and sample holder are still in the conceptual design stage. The sample holder dimensions are similar to the EDIPO design [1]. We are expecting to start finalizing this design after a meeting with users later this year and finalizing the top plate assembly design. The anticryostat is designed to house a sample, a superconducting transformer, and cooling piping (Fig. 6). Between test runs, the sample holder can be removed and reinserted without warming up the magnet cold mass. The anticryostat is designed to operate between 4.5 to 50 K, where the temperature will be controlled per user request. This anticryostat will be made from 316 L stainless steel inner and outer specially profiled tubes. The interspace between the tubes will be under vacuum and partially filled with MLI. For mechanical stability of the anticryostat, supporting spiders will be installed in the vacuum space between the tubes. The spiders will be built from low heat conduction material. The profile of the sample holder is rectangular and will be suspended from the top plate assembly and magnet support plates. The sample holder housing will be made from G-10 for its non-magnetic, conductive, and mechanical properties and will be used to hold the cables at the desired spacing and position inside the magnetic field while withstanding the forces generated under ultimate magnetic flux. Inside the sample holder housing will be a capillary for liquid helium bath distribution and gas evaporation flow during operation. Similarly, the anticryostat for R&D magnets will be similar to the one used currently on the VMTF and the inner space will typically operate under vacuum or room temperature; the design of this will, however, come later in 2023.

IX. CONCLUSION

Fermilab is building a new high-field cable-testing facility with a capability similar to that of the European facilities EDIPO and FRESCA2. It will serve two US national programs within the DOE Office of Science, the Magnet Development Program, and the US Fusion Energy Sciences programs, by making it possible to test HTS samples in a 15 T field.

This paper reports on the progress of the design and construction of the facility. The construction of the pit is finished. The cryostat and the top and lambda plates are approaching final engineering design. It is expected that they will be produced later this year. Two sets of power supplies are under bid. After a workshop with potential users later this year, we will finalize the anticryostat and the sample holder parameters.

REFERENCES

[1] P. Bruzzone *et al.*, "EDIPO: The Test Facility for High-Current High-Field HTS Superconductors," *IEEE Trans. Appl. Supercond.*, vol. 26, no. 2, pp. 35-40, March 2016.
[2] A.P. Verweij *et al*., "1.9 K test facility for the reception of the superconducting cables for the LHC," *IEEE Trans. Appl. Supercond.*, vol. 9, no. 2, pp. 153-156, June 1999.
[3] S. Prestemon *et al*., "The 2020 Updated Roadmaps for the U.S. Magnet Development Program," https://science.osti.gov/hep/Community-Resources/Reports
[4] G. Velev *et al*., "Design and Construction of a High Field Cable Test Facility at Fermilab," *IEEE Trans. Appl. Supercond.*, vol. 31, no. 5, pp. 1-4, Aug. 2021.
[5] J. Fernandez *et al*., "Development of the $Nb_3Sn$ coils for the 15 T cable test facility dipole magnet", Submitted to this conference, 2LOr2A-06.
[6] J. Elen *et al.,* "The superconductor test facility Sultan," *IEEE Trans. on Magnetics*, vol. 17, no. 1, pp. 490-493, Jan. 1981.
[7] S Koshelev *et al*.," Design of the cryostat for High Field Vertical Magnet Testing Facility at Fermilab," 2022 IOP Conf. Ser.: Mater. Sci. Eng. 1240 012081.
[8] T.J. Peterson, R.J. Rabehl and C.D. Sylvester, "A 1400 Liter 1.8K Test Facility," *Adv. Cryog. Eng.* Vol. 43, pp 541-548, 1998.
[9] A. Vande Craen *et al.*,"Thermal performance of the new superfluid helium vertical test cryostats for magnet tests at CERN," AIP Conference Proceedings 1573, 229 (2014),
[10] A. Vande Craen *et al.*," Thermal performance of the new superfluid helium vertical test cryostats for magnet tests at CERN," *IOP Conference Series: Materials Science and Engineering*, vol. 502 (IOP Publishing) p. 012081 (2019).
[11] J. Muratore *et al.*," Design and Fabrication of the 1.9 K Magnet Test Facility at BNL, and Test of the First 4-m-Long MQXF Coil," in *IEEE Transactions on Applied Superconductivity*, vol. 28, no. 3, pp. 1-4, April 2018.
[12] https://en.wikipedia.org/wiki/Nationally_Recognized_Testing_Laboratory
[13] A. Galt *et al*., "A Quench Detection and Monitoring System for Superconducting Magnets at Fermilab," *IEEE Trans. Appl. Supercond.*, vol. 32, no. 6, pp. 1-4, Sept. 2022.
[14] National Instruments CompactRIO, information available: https://www.ni.com/en-us/shop/compactrio.html
[15] E. Ravaioli *et al*., "New, Coupling Loss Induced, Quench Protection System for Superconducting Accelerator Magnets", *IEEE Trans. Appl. Supercond.*, vol. 24, no. 3, pp. 1-5, June 2014.